\documentclass[12pt,preprint]{aastex}

\newcommand{\ha}{H$\alpha$}
\newcommand{\hb}{H$\beta$}
\newcommand{\hc}{H$\gamma$}
\newcommand{\vi}{$V\!-\!I$}
\newcommand{\ri}{$R\!-\!I$}

\newcommand{\har}{$H\alpha\!-\!R$}

\shorttitle{ChaMPlane WIYN Spectroscopy}
\shortauthors{Rogel et al.}

\begin{document}


\title{Three Years of ChaMPlane Northern Field WIYN Spectroscopy} 


\author{A. B. Rogel, P. M. Lugger, H. N. Cohn}
\affil{Department of Astronomy, Swain Hall West 319, Indiana University, 727 E. 3rd. St., Bloomington, IN 47405}
\email{abrogel@astro.indiana.edu}

\author{S. D. Slavin}
\affil{Department of Chemistry and Physics, Purdue University Calumet, 2200 169th St., Hammond, IN 46323-2094}

\author{J. E. Grindlay, P. Zhao, J. Hong}
\affil{Harvard-Smithsonian Center for Astrophysics, 60 Garden St., Cambridge, MA 02138}


\begin{abstract}

We present initial results of WIYN spectroscopic observations of
selected objects detected in the \emph{Chandra} Multiwavelength Plane
(ChaMPlane) Survey
in fields towards the Galactic anti-center.  ChaMPlane is designed
to identify low luminosity X-ray sources, both accretion-powered and 
stellar coronal, in the Galaxy.  It also includes a wider-fields optical
imaging Survey conducted with the NOAO Mosaic cameras to identify optical
counterparts as well as \ha-selected objects in the $\sim5\times$ larger field.
We report spectra for 1048 objects in galactic anti-center
(i.e.\ northern) fields, resulting in 609 type determinations.  These
include 5 new cataclysmic variables, 4 Be stars, 14 lithium-absorption
stars, 180 stellar coronal sources (primarily dMe stars), 
and 29 new quasars.  Bright optical counterparts of \emph{Chandra} 
sources in this sample are most
frequently dMe stars whereas a majority of the faintest (R $\sim21$) 
spectroscopically classified \emph{Chandra} source counterparts are
quasars.  The bulk of \ha-selected sources appears to
be roughly evenly divided between dMe stars and M stars at all
magnitudes.  
\end{abstract}




\section{Introduction}

The ChaMPlane Survey \citep{gri2003,gri2004} is designed to detect and
optically identify low luminosity X-ray point sources in the Galactic
plane. ChaMPlane is conducted in two phases: the ChaMPlane X-ray
survey of serendipitous sources in \emph{Chandra} X-ray Observatory
archival deep ($>$20 ksec) Galactic plane ($|b|<12^\circ$) fields, and
the ChaMPlane Optical Survey of deep $V, R, I,$ and $H\alpha$ images of
fields taken at CTIO or KPNO with the 4m telescope Mosaic $(36'\times36')$ cameras
centered on the \emph{Chandra} fields.  Detailed descriptions of the
X-ray detections and database are given by \citet{hon2004},
and the optical imaging, astrometry, and photometry with
Mosaic are described by \citet{zha2004}.

The ChaMPlane X-ray Survey 
takes advantage of the
unparalleled angular resolution and $\sim1''$ absolute astrometry
of the \emph{Chandra} X-ray Observatory to match the serendipitous
X-ray sources in the field with faint ($R < 24$) optical counterparts.
In all but the most crowded 
fields, nearly all matches are unambiguous.  The point-source
sensitivity for a typical 30 ksec ChaMPlane exposure allows detection
of sources with $L_X \gtrsim 10^{31}~{\rm erg}\,{\rm s}^{-1}$ at 8 kpc,
which enables detection of cataclysmic variables (CVs), one of the
prime ChaMPlane objectives, throughout a significant fraction of the
Galaxy.  \citet{gri2003} and \citet{zha2003} have reported early
results from the ChaMPlane X-ray Survey and followup optical
photometry.

The ChaMPlane Optical Survey provides optical photometric followup of
the \emph{Chandra} serendipitous sources, providing \har, \ri, and
\vi\ colors for the X-ray sources.  By definition, stars without \ha\
emission should have
an average \har\ color index of $\sim$0.0, while CVs and other
accretion-powered systems frequently have significant \ha\ excesses,
which register as negative \har\ colors.  This enables us to select
objects that are \ha\ and/or X-ray bright for followup spectroscopy
and classification.

The ChaMPlane Optical Survey will include all \emph{Chandra} fields (observed
with either the ACIS-I or ACIS-S cameras, with $16' \times 16'$ and
$8' \times 8'$ fields, respectively) in the ChaMPlane X-ray
Survey. The highest priority for the spectroscopic followup is to
obtain spectra on each X-ray source optical counterpart with possible
\ha-emission; the second priority is to obtain spectra on {\it all}
\emph{Chandra} source optical counterpart candidates, regardless of
\ha-emission status.  Since the Mosaic camera has a field of view
about 4.5 times larger in area than the ACIS-I, the Optical Survey will
also detect many \ha-emission sources with unknown X-ray emission
status.  Thus, as a third priority, spectroscopy is also 
performed on these objects, since Hydra can simultaneously observe
60--70 objects, which (for the anti-center fields at least) enables us
to observe all \ha-emission candidates in the ChaMPlane Optical Survey
field in just one or two Hydra setups.

The ChaMPlane Optical imaging program is described in detail by \citet{zha2004}.  
Normally, three (for $V$ and $I$) or five (for $R$ and
\ha) dithered exposures are taken and combined to remove cosmic rays
and other image contamination.  Total exposure times are set to enable
5--7\% precision photometry to 24th magnitude in all bands.  Short exposures in
each band are also taken to give unsaturated data for the brighter
objects.  Mosaic data reduction and analysis is done using standard
IRAF\footnote{IRAF in distributed by the National
Optical Astronomy Observatories, which are operated by the Association
of Universities for Research in Astronomy, Inc., under cooperative
agreement with the National Science Foundation.}
\emph{mscred} tasks, and photometry with DAOPHOT.

In this paper, we report initial results for followup spectroscopy carried out
with the WIYN\footnote{The WIYN Observatory is a joint facility of the
University of Wisconsin, Indiana University, Yale University, and the
National Optical Astronomy Observatories.} Hydra multi-object
spectrometer.
This paper is organized as follows: \S2 describes the observing and
analysis precedures, with spectroscopic results presented in
\S3.  We summarize our results in \S4.
In \citet{rog2004}, we discuss some of the implications of 
our anti-center field results
for the space density of CVs, using a Monte-Carlo model of 
the CV distribution within the Galaxy.
\citet{lay2004}
discuss the combined X-ray and optical (both 
photometric and spectroscopic) constraints on the total \emph{Chandra} 
source populations in the anti-center fields, as well as X-ray properties 
of individual classes of objects.

\section{Data Acquisition and Reduction}
%%%
\subsection{Field Selection}

\begin{figure}
\plotone{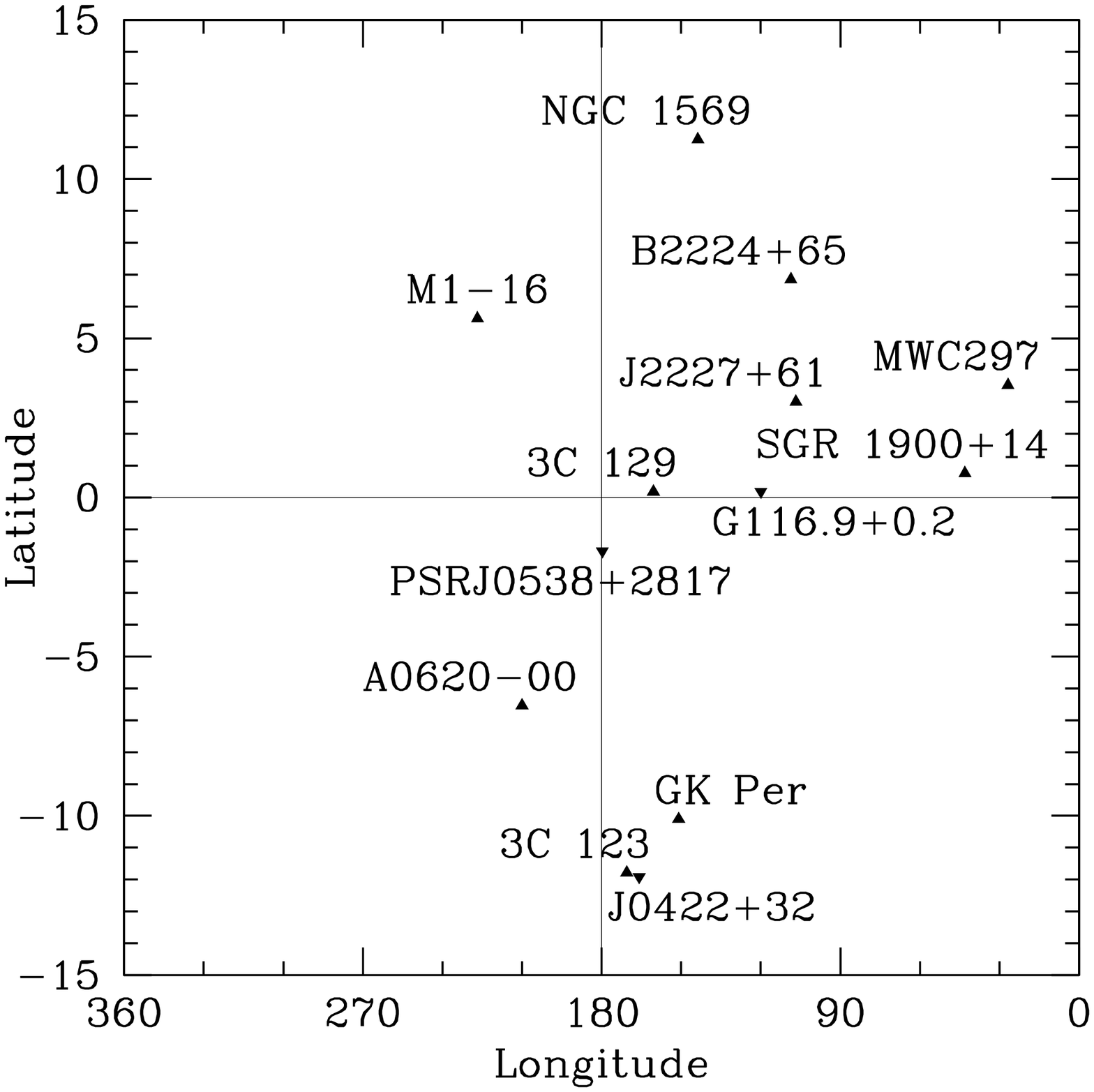}
\caption{Field positions in Galactic coordinates.  The actual field
positions for J0422+32, PSRJ0538+32, and G116.9+0.2 are above the names,
 the rest are below the
names. \label{fields}}
\end{figure}

Brief summaries of the 13 fields observed with WIYN/Hydra to date are given in Table
\ref{tbl-fielddata}.
Of these 13 fields, 11 are in the general Galactic
anti-center direction (quadrants 2 and 3, i.e. galactic longitude
$90^\circ \le l \le 270^\circ$), as seen in Figure \ref{fields}. 
Eight of the fields lie within $45^\circ$ of the
Galactic anticenter.
The two
fields in the first quadrant (Sgr1900+14 and MWC297) are 
included since their northern declination allowed them to be observed
from WIYN\@.  
These two first quadrant fields still sample pure
disk volumes, as the Galactic bulge extends only to about 3\,kpc
\citep{bin1998}, which corresponds to about $l=22^\circ$.  An
extensive program of photometry and spectroscopy on Galactic bulge fields
is also being conducted by ChaMPlane and will be reported in subsequent
papers.

\subsection{Spectroscopic Target Selection}

Each \emph{Chandra} field was observed with the CTIO or KPNO 4m in the $V$,
$R$, $I$, and \ha\ bands with the Mosaic ($36' \times 36'$) camera.
Target lists were created for both X-ray and \ha\ objects.  The X-ray
target list is composed of objects detected in the \emph{Chandra}
observation which have optical counterparts in the combined $R$-band image, which
is the deepest of the four bands.  An
X-ray source is considered to match an optical source if 
the position offset is less than a variable offset allowance,
derived from the
optical astrometry error ($<1''$), X-ray--optical boresite error 
($<1''$), and the X-ray astrometry ($\sim0.3''$
for the best case of a bright source near the instrument axis, and up to $4''$ for faint
objects far from the instrument axis).  The maximum offset allowance is $4''$, but
sources near the ACIS axis have offset allowances as small as $1''$.
Roughly 50\% of \emph{Chandra} sources in these
13 anti-center fields have optical counterparts with $R < 23$ \citep{zha2003}.
Details of the astrometry and optical vs. X-ray matches 
are given by \citet{zha2004} and \citet{lay2004}.

The \ha-definite target list is composed of objects selected by
\har\ color, with \har\ $\leq-0.3$ defined as the minimum threshold for objects of definite interest.
Given the 
KPNO MOSAIC filter bandpass data for the \ha\ and R filters, 
our threshold corresponds to EW(\ha) $\sim 28$ \AA, using
a flat continuum spectrum with \ha\ emission added at 6563\AA.
This criterion allows us to reject most normal stars, while retaining potential
strong \ha -emission CV candidates.
However, many normal M and dMe stars are
selected by this criterion, as a result of the
strong TiO absorption bands on either side of  $\sim$6500 \AA\@.
A separate \ha\ target list (the \ha-marginal, or \ha M, list) is composed of objects with \har\
color between $-0.2$ and $-0.3$, or EW(\ha) between 17 \AA\ and 28 \AA.  
These targets are of lower interest, 
since the object is more likely to be a normal star due to the spread in the 
\har\ distribution.

\emph{Chandra} sources which are also \ha\ targets receive highest priority 
for the assignment of Hydra fibers, as these are the most promising X-ray binary or 
CV candidates.
All remaining \emph{Chandra} sources with optical counterparts 
with $R < 22$, and thus accessible to WIYN/Hydra, are
given second priority for fiber assignment.  Third in priority are 
\ha\ sources with no \emph{Chandra} detection, usually due to their being 
outside the \emph{Chandra} field of view, which is 5$\times$ smaller in area than the
Mosaic image field.  However, some of these objects are within the 
\emph{Chandra} field but are not 
detected owing to X-ray flux emission below 
$L_X=10^{31}~{\rm erg}\,{\rm s}^{-1}$.
Lowest priority is 
assigned to the \ha-marginal targets.  Before final assignment, the
Mosaic image positions for each selected Hydra target are manually 
verified to check for source confusion, cosmic ray 
contamination, and other problems.

ChaMPlane source naming  follows the following 
convention \citep[cf.][]{gri2004}:  
all X-ray sources are 
designated with a CXOPS (Chandra X-ray Observatory Plane 
Survey) prefix, followed by
J2000 X-ray coordinates JHHMMSS.S+DDMMSS.  All
optical sources are designated with a ChOPS (Chandra 
Optical Plane Survey) prefix, followed by optical coordinates 
JHHMMSS.SS+DDMMSS.S. 
For an X-ray source with an optical counterpart, the
CXOPS and ChOPS coordinates can differ by up to the offset allowance 
discussed above 
owing to the different sources for the coordinates.  
As all X-ray sources selected for 
follow-up spectroscopy have optical counterparts, they will
have both CXOPS and ChOPS designations.  
In this paper, only the CXOPS
designation is used for objects with \emph{Chandra} detections.  Objects 
referred to by a ChOPS designation in this paper 
include \ha-emission objects selected for spectroscopy but for 
which there 
is either a \emph{Chandra} non-detection or no \emph{Chandra} measurement 
at all.
Both the CXOPS and ChOPS catalogs cross--reference each other and 
be released in future papers.

\subsection{Spectroscopic Observations}
All the spectra presented here were taken with the Hydra fiber-fed
multi-object spectrograph on the 3.5m WIYN telescope.  Hydra has two
fiber bundles, with different-sized fibers, optimized to work at
different wavelength ranges.  We chose to use the red bundle, which
has smaller fibers (2\arcsec\ diameter versus 3\arcsec\ for the blue fibers), 
to reduce the impact of skylight on the spectra.
This, along with the bench spectrograph's CCD sensitivity profile, results in severely
diminished blue sensitivity.  As we are primarily interested in \ha, at
6563 \AA, and nearby helium lines, this is an acceptable
compromise.

Field setup is done with the \emph{whydra} program. 
ChaMPlane objects are assigned to fibers with the priorities given above.
Each
sublist is sorted by R-band magnitude to give preference to bright
objects.  About 15-25 sky positions are used, as well as a minimum of 6 FOPS
(guide stars), leaving 65-75 object fibers available.  The \emph{whydra}
output is then tuned by hand using the \emph{hydrasim} package to optimize
fiber usage, resulting in about 65 objects per field setup.

We use the $600@10.1$ ``Zepf" blue-blazed grating for Hydra, which
results in a resolution of 1\farcs4/pixel.  
We target a wavelength range of 4000--6800 \AA, which gives us good coverage
of the first few Balmer lines and the helium lines at 4686 \AA, 5875 \AA, and 6678 \AA,
while avoiding the extremely bright comparison-lamp line at 6965 \AA.  This enables
us to take a single comparison exposure with good S/N over the entire wavelength
range.  The typical observing procedure is to
take multiple sets of two long ($2400-3600$ sec) exposures, with each set 
bracketed by 200 sec CuAr comparison lamp exposures.  This minimizes the time lost 
to calibration, while still allowing for correction of wavelength drift over
the night.  To minimize the effect
of cosmic rays, at least 5 exposures are taken, with the available time split
to allow for this.  This results in a total of
3.5--7 hours of exposure per setup.  Some fields have enough objects that multiple
setups are necessary, which allows for longer total exposure time on
fainter objects and objects of particular interest.  Depending on the
season, 1--2 setups per night are observed.

\subsection{Reduction Procedure}
The basic data reduction process consists of preliminary image
processing using the IRAF
\emph{ccdproc}
package (for bias subtraction), followed by spectral reduction using
the IRAF \emph{dohydra} package.  Standard procedures were followed,
with dome flats used both for aperture reference and flat fields.  Sky
editing was used to remove problematic sky fibers.  Bracketing
comparison spectra were always used in the reduction process.
Combination of the resulting spectra was done with a median filter to
eliminate cosmic rays.

For the B2224, J2227, G116, MWC297, and PSRJ0538 fields, the \ha\
image showed substantial variations in the background signal,
resulting from Galactic diffuse emission in \ha, \ion{N}{2}, and \ion{S}{2}.  
For the
case of the fields with this diffuse emission, the basic reduction
process results in the subtraction of an \emph{average} (sky +
Galactic diffuse) emission level from all of the spectra.  As the
Galactic diffuse emission level varied across the field, this process
generally results in over- or under-subtraction of the diffuse
emission in each individual spectrum.  As we are specifically
interested in the \ha\ emission levels of the targets, this can be a
serious problem, as became clear from the first observation of the
B2224 and J2227 fields in August 2002.
We therefore remove the Galactic diffuse emission signal by taking
spectra of sky positions offset by a few arcseconds from the target
object.  Offset positions were selected by careful inspection of the
deep $R$ image, in order to maximize faint object detection to ensure 
clear sky locations.  
In addition, the deep \ha\ frame was used to pick a location
for the offset fiber that had the same (or as close as possible) 
Galactic diffuse emission as the object location.  
This proved especially important
in the few cases where there was a supernova remnant in the Mosaic image.
Fibers originally targeted at sky positions were not
moved and thus provide a measure of any change in the sky brightness
between the ``on'' and ``off'' exposures.    

The following outlines the general method used to recover the object
spectrum from the observations, with actual observations indicated
with parentheses.

\begin{eqnarray*}
\mbox{``GDE''}& = & \mbox{Galactic Diffuse Emission} \\
\mbox{``Sky''}& = & average{\mbox{(GDE + local sky) for sky-fiber positions}} \\
\mbox{``On''} & = & (\mbox{Object} + \mbox{GDE} + \mbox{local sky}) - \mbox{Sky} \\
\mbox{``Off''}& = & (\mbox{GDE} + \mbox{local sky}) - \mbox{Sky} \\
\mbox{``Object''} &= & \mbox{On} - \mbox{Off}
\end{eqnarray*}

Typically, about 1 hour of observing time is devoted to the offset
positions, split into 5 separate observations to enable good cosmic
ray removal.  
As a consistency test, the 
residual line emission in the sky fibers (which are not moved between the
``on'' and ``off'' setups) is checked to ensure that it 
falls within the general expected noise level.
This method
has been applied with good success since the 2002 Dec 5 run.

In cases where we experienced poor weather during a run in which this
method was used, 
we simply looked at the line ratios in the
diffuse Galactic emission, versus the same line ratios in the object
spectrum.  In several cases, the \ha\ vs.\ [\ion{N}{2}]$\lambda$ 6548.1 \AA\ or
[\ion{S}{2}]$\lambda$ 6716.4 \AA\ and $\lambda$ 6730.8 \AA\ ratios have been substantially
higher or lower in the object than in the offset spectrum, indicating
clear \ha\ emission or absorption in the object.  Subsequent
application of the offset method during a later run has supported the
positive emission classifications of this ratio method.

Classification of objects is done by visual inspection of reduced
spectra.  For CVs, we expect to see strong Balmer emission, possibly with 
split profiles, \ion{He}{1} emission, and possibly \ion{He}{2} emission.  Other objects,
such as T Tauri stars and M dwarf flare stars, can have both Balmer and \ion{He}{1}
emission, but usually with (much) narrower lines.  
Thus, additional criteria are needed to distinguish between these
and likely CV candidates.  The presence of lithium absorption is used to 
distinguish T Tauri (or other young star) candidates, and the FWHM of the
\ha\ line is used to determine flare-star candidacy.  Our narrowest line CV candidate,
which shows a split \ha\ profile, has a FWHM of 10 \AA.  The one flare-star
candidate (with both Balmer and \ion{He}{1} emission) has an \ha\ FWHM of 4.5 \AA.
Comparison with published spectra was used for spectral
classification of stars.  Quasar redshifts were determined by comparing
emission line wavelengths to those given by \citet{vanb2001}.

\section{Spectral Results}	
\subsection{Overview}
\begin{figure}
\plotone{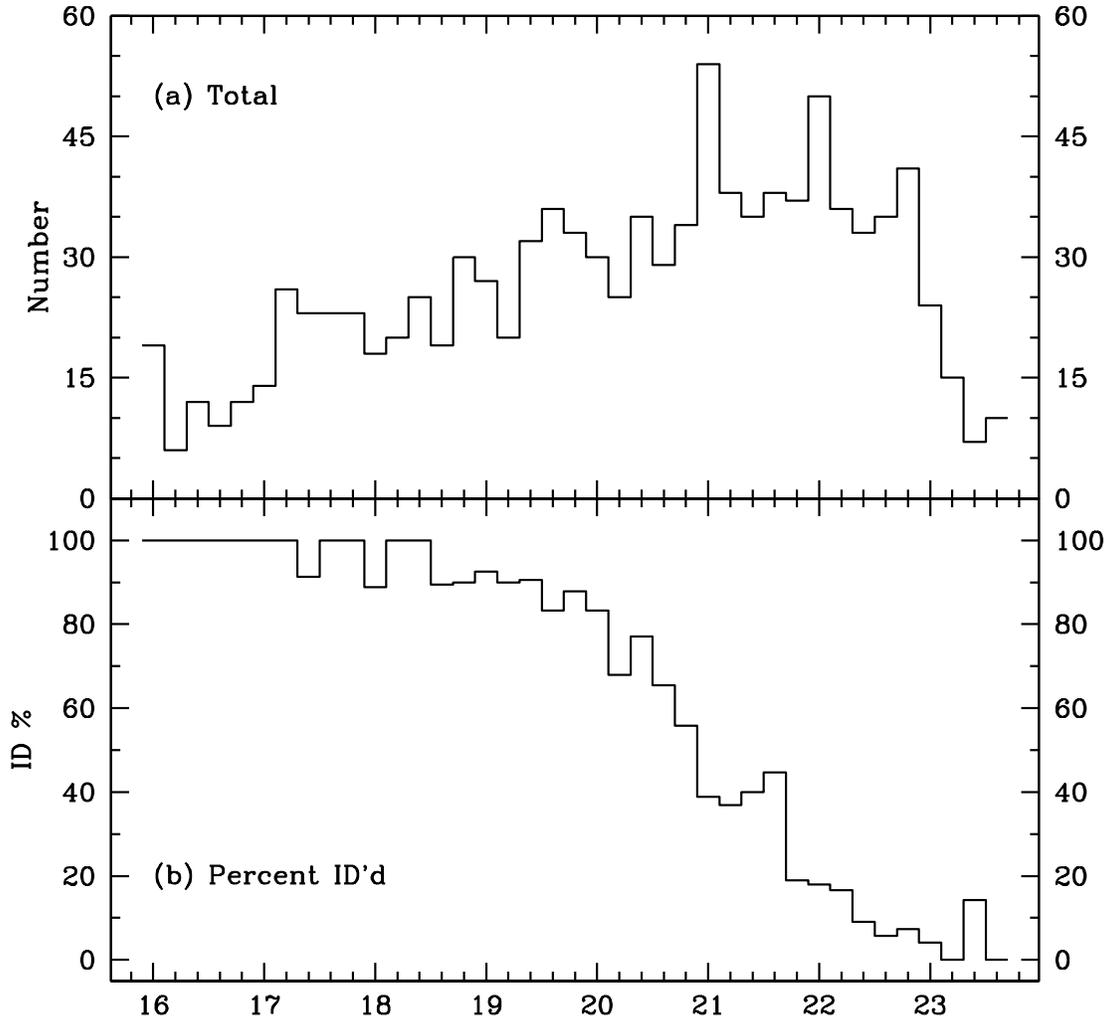}
\caption{Histograms of R-band magnitude versus total observations (panel a) and 
classification rate (panel b).  \label{histor}}
\end{figure}

To date, we have observed 13 fields over 17 nights, 
with multiple setups on some 
fields.  A summary of the observing runs is 
given in Table \ref{tbl-fieldepochs}.
Table \ref{tbl-obssum} lists the total, observed, and classified number of
objects for each target list, summed across all fields.  
Five new CV candidates were detected within the \ha-definite list.
Of the 14 objects designated as `duplicates',
seven appear on both the X-ray and \ha-definite list, and seven appear on 
both the 
X-ray and \ha-marginal list.  
The `other' list is a set of objects that
were observed early in the spectroscopic program, but that were subsequently 
dropped from the target lists as a result of refinements of the targeting criteria.
In general, about 50\% of targets can 
be classified, with classification failure primarily due to the WIYN 
magnitude limit.  
The \ha-marginal list has a much higher 
classification rate, owing to selection of the brighter marginal 
objects to fill out a field.  
Figure \ref{histor} shows the 
classification rate as a function of target magnitude, demonstrating 
that we have near-complete classification of objects brighter than $R=20$.
We can generally classify about half of the objects as faint as $R=21.5$, and
in some cases can classify
objects as faint as magnitude 23.5.  These latter objects are typically those 
with very strong emission characteristics, such as narrow-line
quasars.  

\begin{figure}
\plotone{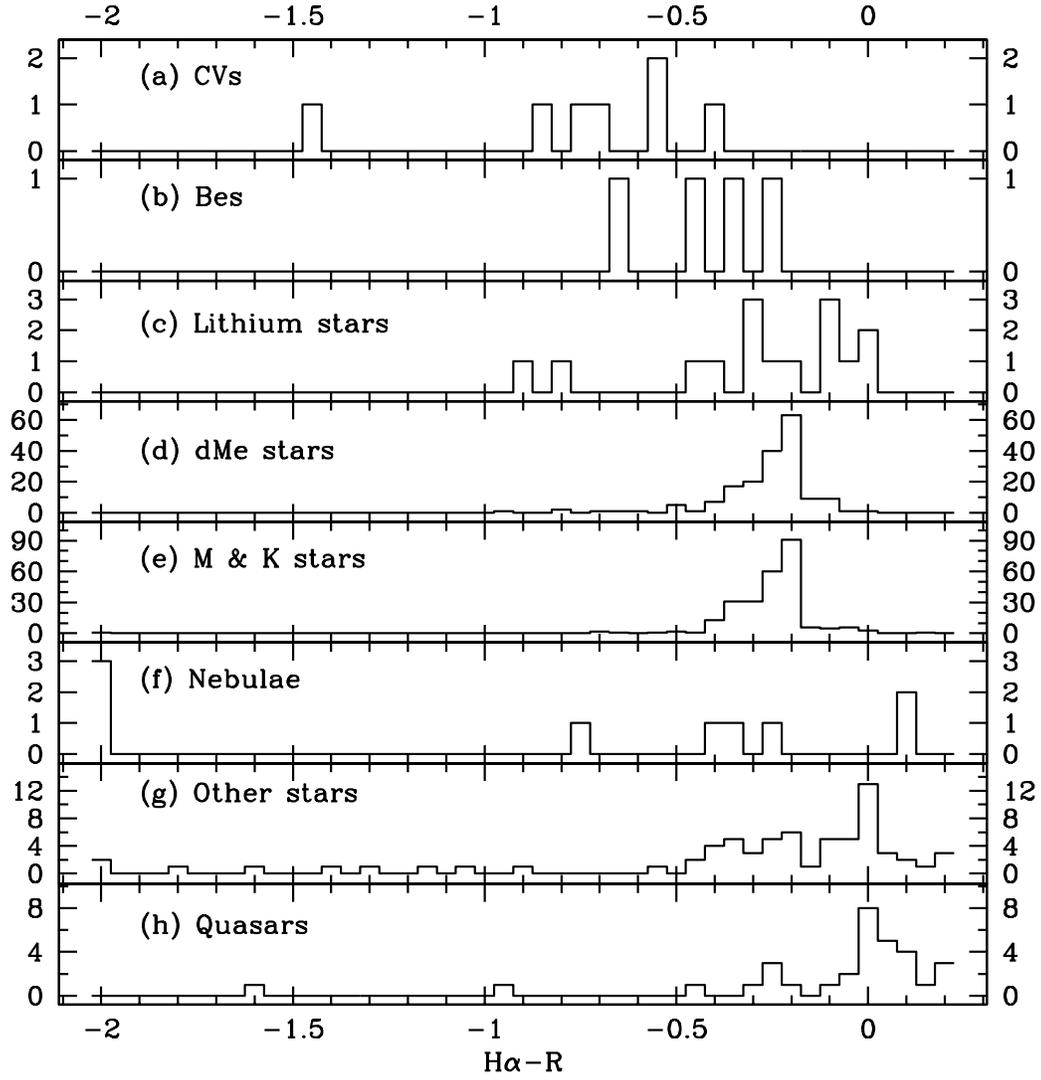}
\caption{Histograms of \har\ color of various object types detected in the
project.  The \har\ color cutoff for non-X-ray objects of definite interest
was at $-0.3$; the cutoff for objects of marginal interest was at $-0.2$.  
\label{histohr}}
\end{figure}

\begin{figure}
\plotone{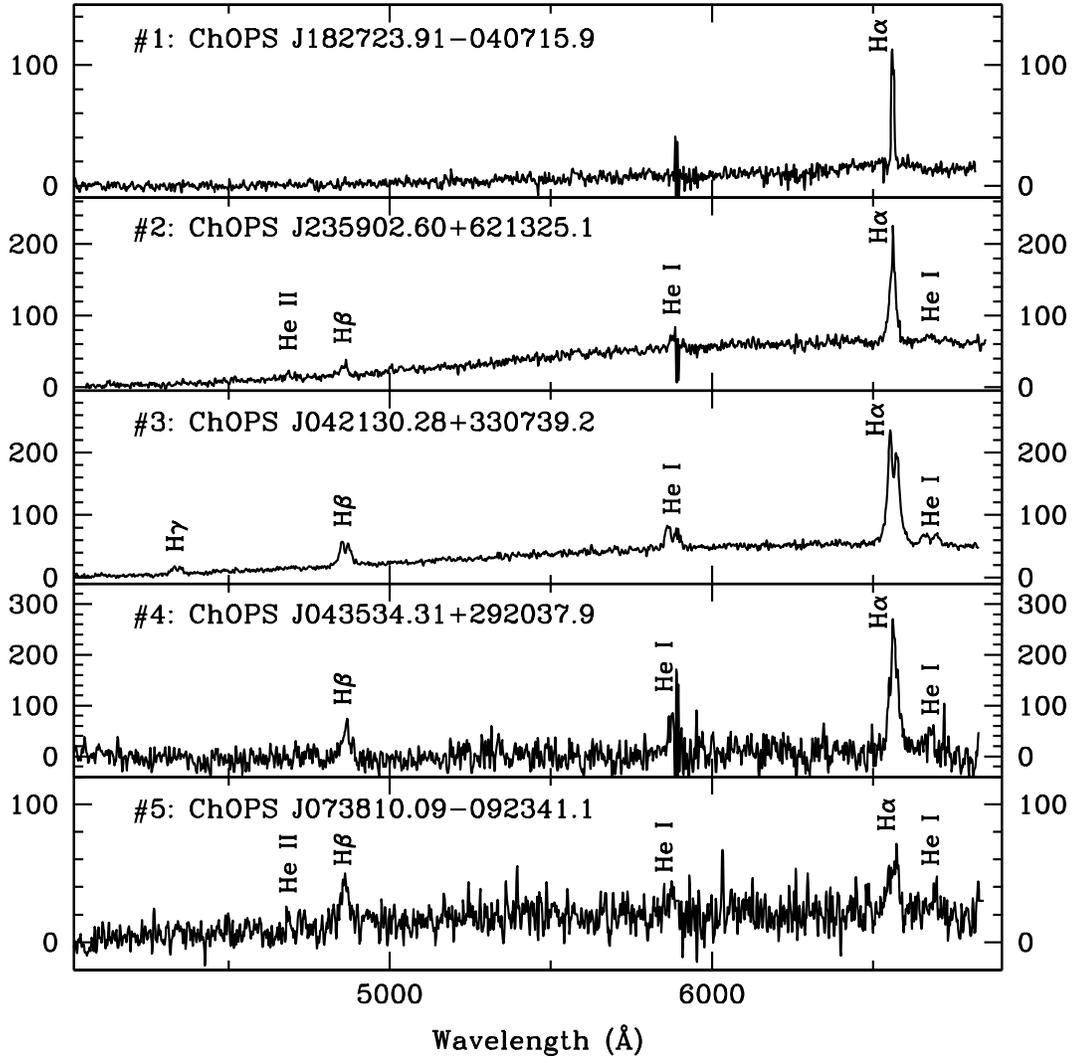}
\caption{All new CV candidates detected to date, smoothed with a 3-channel boxcar filter. \label{allnewcv}}
\end{figure}

\begin{figure}
\plotone{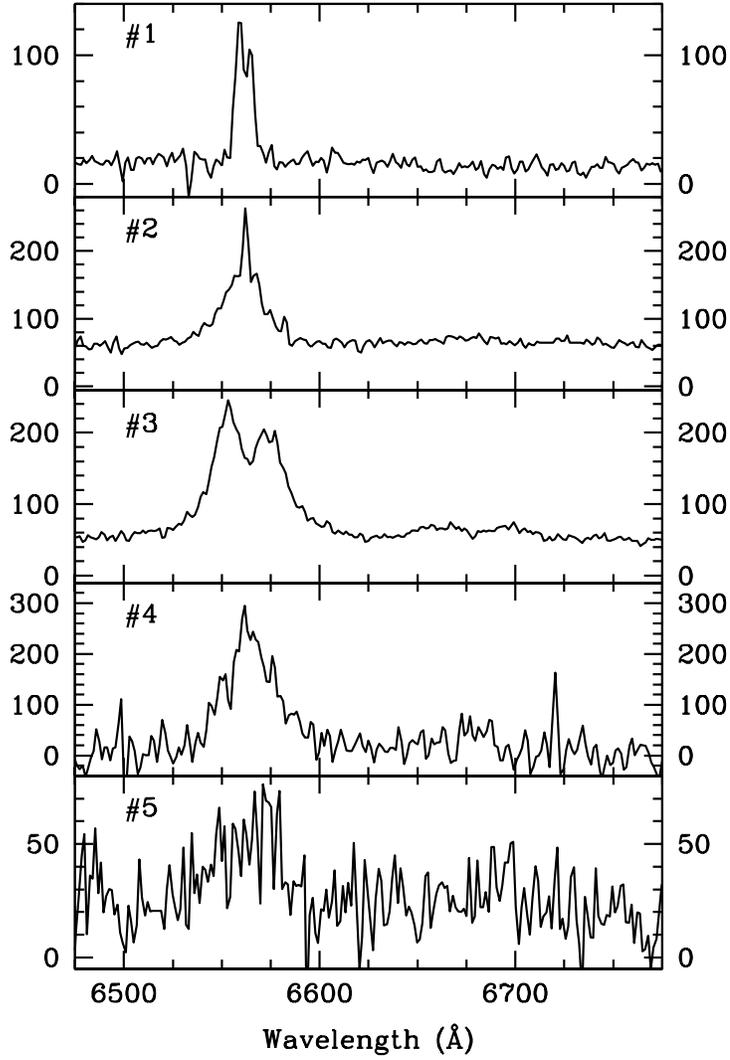}
\caption{Unsmoothed \ha\ and \ion{He}{1} emission regions of all new CV candidate objects. \label{allnewcvha}}
\end{figure}

As a check of our \har\ cutoffs for \ha-definite and \ha-marginal targets,
we looked at object classifications binned by \har\ color 
(Figure \ref{histohr}).
All five CV candidates, the two known LMXBs, and three of four Be
stars are found to have \har\ color $\leq-0.3$, and all the CV/LMXBs but A0620-00
have \har\ color $\leq-0.5$.  
Only one Be star is found with an \har\
color between $-0.3$ and $-0.2$. 
Nearly all M and K stars of all types have measured \har\ color $\geq-0.4$, with most
being $\geq-0.3$.
Thus, with the placement of the \har\ cutoffs, we achieve our goal of limiting 
the selection of typical stars while not excluding our objects of prime interest.
As expected, our quasars 
usually have a near-zero apparent \har\ color, 
as only certain specific values of redshift will cause
a typical quasar or AGN emission line
to be located within the \ha\ bandpass.  
There appears to be no correlation between the
photometrically-determined \har\ color and the
spectroscopically-determined emission status for M/K stars. 
As the \ha\ emission of dMe stars is known to vary by a factor of 2 on a timescale of
a day \citep{rei2000}, and these objects are also flaring objects, 
this lack of a correlation is not unexpected.
Variability could also explain why the Be star in the 3C 129 field  
was on the \ha-marginal target 
list.
Spectra for our newly-discovered CV candidates and a representative sample
of the other interesting object classes are given as Figures 
\ref{allnewcv}-\ref{quasar}.  
As we are primarily interested in simply classifying
these objects, all spectra shown are NOT flux-calibrated.

\subsection{CVs}
Figure \ref{allnewcv} shows the five CV candidates
(all outside the \emph{Chandra} field of view, but in the Mosaic field)
discovered to date using WIYN/Hydra, while Table \ref{tbl-cvs} lists the
reference numbers and basic parameters for all five CV candidates and two 
previously-known LMXBs observed so far.  
Figure \ref{allnewcvha} shows the \ha\ and \ion{He}{1} $\lambda$6678 \AA\ regions 
for all new CV candidates.

CV \#3 
exhibits broad, split-profile 
\ha\ and \ion{He}{1} emission lines, making
it the clearest candidate for CV status.  The \ha\ line split gives a $v\,\sin\,i$ rotation
of 480 km/s.  

CV \#2  
and CV \#4  
both have broad
Balmer and \ion{He}{1} lines, without showing the split profile.  However, Galactic
diffuse emission is present in CV \#2, causing the narrow spike in the center of the \ha\ 
emission line and potentially filling a split line, and the 
CV \#4 spectrum is noisy, so 
a split profile could be masked in both cases.    
CV \#2 also exhibits the higher-excitation line \ion{He}{2} $\lambda$4686 \AA. 

CV \#5 
is similar to CV \#2, with very weak evidence for the \ion{He}{2} $\lambda$4686 \AA\ line.  
The spectrum is quite noisy, as we only have one hour of data.  
Acquisition of more data on this object is planned for Fall 2004.

CV \#1 only shows a narrow
\ha-emission feature, but it has a resolved, split peak characteristic
of an accretion disk, and the signal is low enough that the other
lines might be buried in the noise.  The narrow \ha\ line split gives a
$v\,\sin\,i$ rotation of just 130 km/s, which, if one assumes
rotational properties similar to CV \#3, gives a maximum inclination
of just $15^\circ$.  Comparing the weak, mid-M star continuum profile
of this source to other M stars in the same field gives an estimated
magnitude for the secondary of $R=21.7\pm0.2$.  Using the absolute
magnitudes for M dwarfs given by \citet{thep1984}, we get a distance
of $1200\pm600\,$pc, assuming a secondary spectral type of
between M2 and M6 and a constant extinction in $R$ of 0.97 magnitudes
per kpc \citep{bin1998}.

\subsection{Be stars}
\begin{figure}
\plotone{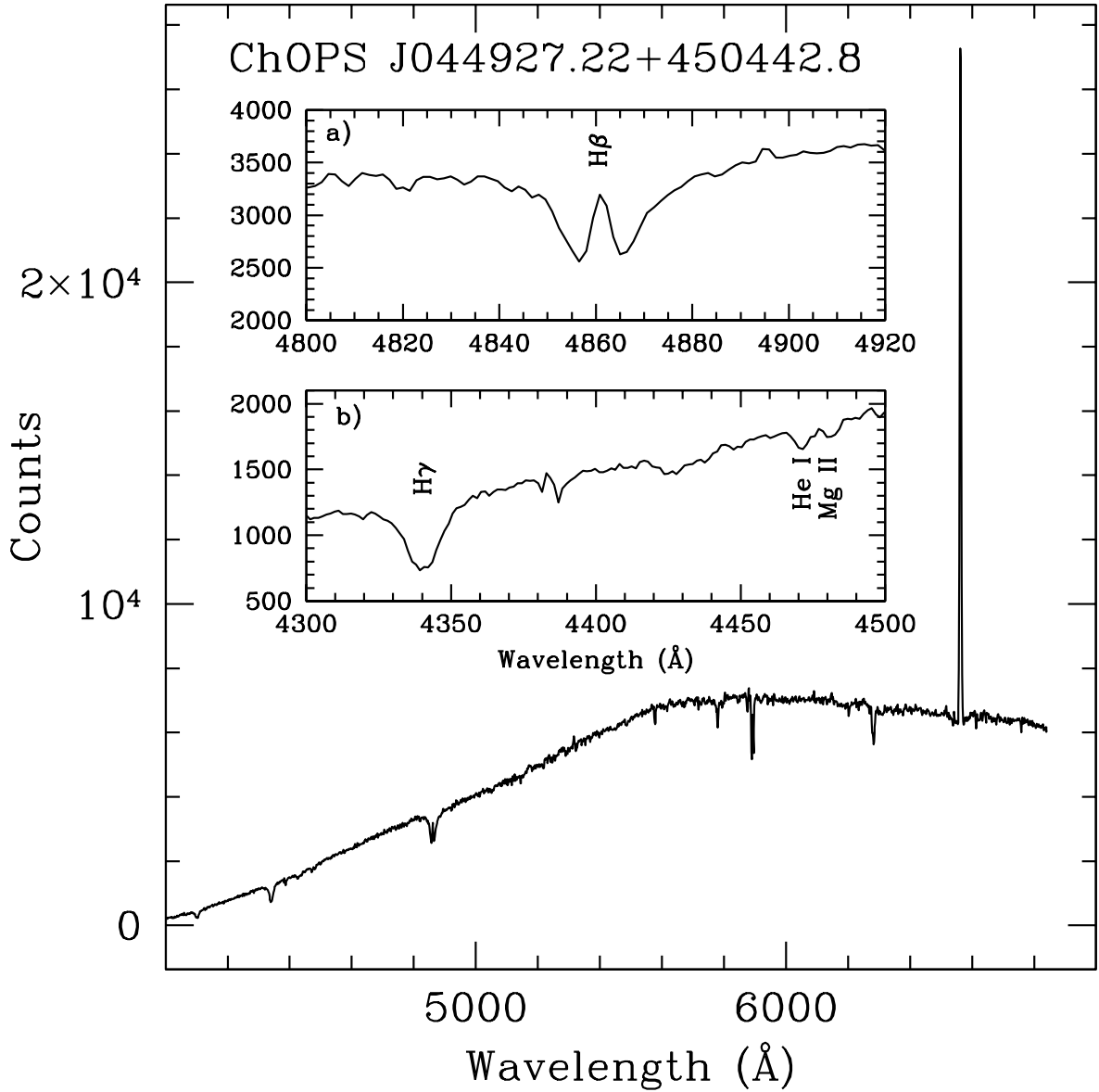}
\caption{
Candidate Be star ChOPS J044927.22+450443.8 in the 3C 129 field.  
Inset panel a) is the $H\beta$ line,
while b) is the $H\gamma$ line and neighboring \ion{He}{1} and \ion{Mg}{2} lines
($\lambda$4471 \AA\ and $\lambda$4481 \AA). \label{bstarnest}}
\end{figure}

Table \ref{tbl-bes} gives the basic parameters of the four Be star candidates classified with our
WIYN spectra.  While two of these objects were located within the ACIS 
field of view, no X-ray detections were made of these targets, thus 
eliminating them as HMXB candidates.  
All are spectrally similar to 
the `classical' Be star of \citet{por2003},
which is a rapidly-rotating massive (late O to early A) star which creates an 
equatorial disk in some
as yet poorly-understood manner.
Figure \ref{bstarnest} shows an example Be star candidate spectra.
Of note in the spectrum 
is the apparently split $H\beta$
line (Figure \ref{bstarnest}, panel a), which is the result of partial
filling of the photospheric absorption line by the envelope emission.
Given the relative strengths of the three lines ($H\gamma$, \ion{He}{1} $\lambda$4471 \AA, and
\ion{Mg}{2} $\lambda$4481 \AA), as seen in inset panel (b) of Figure \ref{bstarnest}, 
it appears that this star is likely to be spectral type B5
\citep*{ste1999}.  
Assuming the constant extinction of 0.97 mag/kpc in $R$ as above, and 
using estimates for absolute
magnitudes \citep{hipp1997} and color \citep{wink1997} from the literature, 
we derive a distance of
$2300\pm300\,$pc.  The other three Be stars are fainter, and therefore have
computed distances of 4--6 kpc.  These distances are actually underestimates, 
since the Be stars are 
located in fields in the second quadrant, i.e. outside the solar circle, and thus the
extinction is most likely lower than assumed.  The
detection of Be stars at only the bright end of our available magnitude range, in our mostly 
quadrant two and three fields, is a result of
a lack of star formation regions in the far reaches of the Galactic disk.  

\subsection{Lithium absorption stars}

\begin{figure}
\plotone{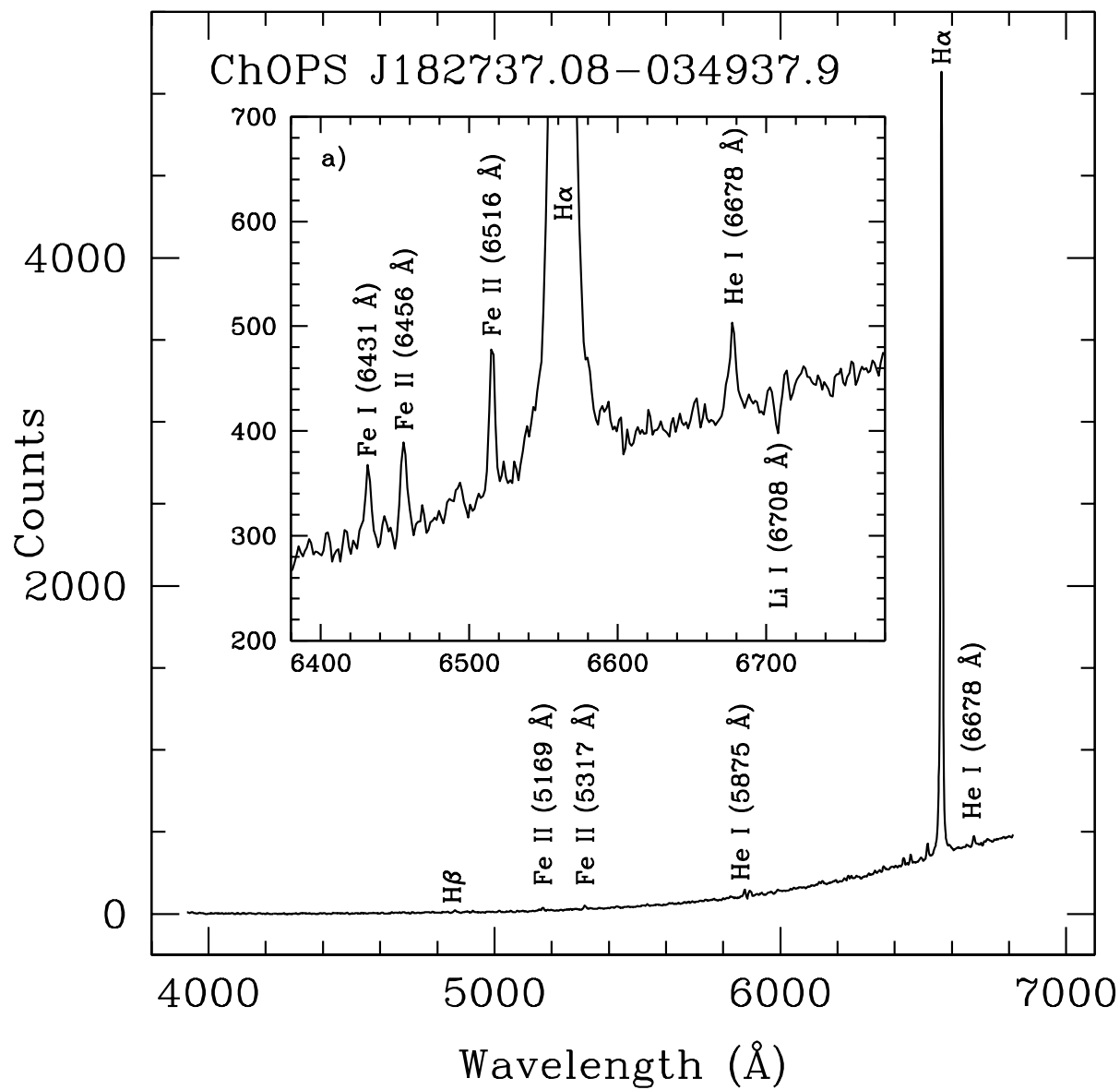}
\caption{New T Tauri candidate object ChOPS J182737.08-034937.9.  
Inset panel a) shows the \ha\ region of the spectrum, showing
iron emission and lithium absorption lines as well as \ha\ and \ion{He}{1} emission.
\label{ttau}}
\end{figure}

\begin{figure}
\plotone{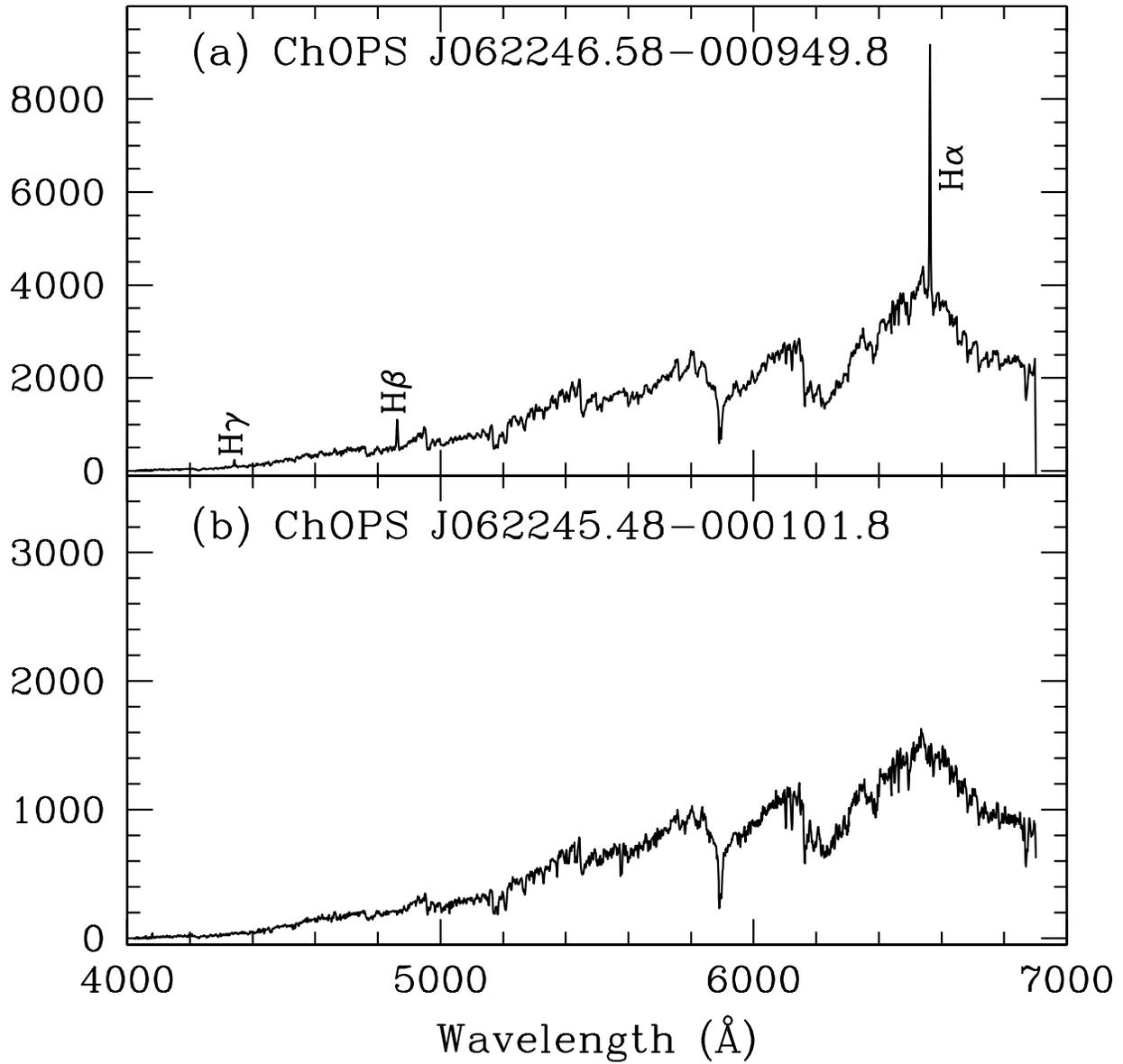}
\caption{dM3e (panel a) and M3 star (panel b) spectra (from A0620 field). \label{mstar}}
\end{figure}

A total of 15 objects show clear lithium 
absorption with equivalent widths of 0.2--0.6 \AA.    
Table \ref{tbl-li} lists their parameters.  
Other than the lithium-abundant LXMB A0620-00\citep{gon2004}, 
these stars are
all simply classified as Lithium stars.
Some show additional
spectral features (as in Figure \ref{ttau}) allowing for further classification
(most likely as T Tauri stars), 
while others show normal stellar continuum profiles, potentially being either very 
young main sequence stars, carbon stars \citep[e.g.][]{hat2003}, or late-type 
giants\citep[e.g.][]{dra2002}.  Of the 10 Lithium stars located within the
\emph{Chandra} field of view, 8 were detected as X-ray sources.  Further discussion 
of the X-ray properties of these objects will be included in \citet{lay2004}.

Figure \ref{ttau} shows an example of a spectrum showing strong clear lithium 
absorption, along with many strong emission features.  Figure \ref{ttau}, inset panel (a)
shows the spectrum from 6400--6800 \AA, which shows several iron emission
lines as well as the lithium absorption.
It is likely that this object is a T Tauri 
star, based on the similarities exhibited between it and the spectrum of DR Tau
\citep{smi1999}.  

\subsection{Coronal emission stars}
A total of 433 M and K stars have been classified to date, with 180 being
classified as clear emission objects.  At present, we 
classify an object as an \ha-emission 
object if it has clearly detectable \ha\ emission when the spectrum
is examined by eye. 
A further 40 objects have possible \ha\
emission, but contamination by diffuse Galactic emission precludes definitive 
classification.
A pair of sample spectra is given in
Figure \ref{mstar}.  Both stars are of similar spectral type \citep[about M3, 
based on Fig. 2.2 in][]{rei2000}, but \ha\ emission
of equivalent width $\sim$5 \AA\ is present in ChOPS J062246.58-000949.8.  \hb\ and \hc\ are also
present, with similar equivalent widths.  
One object of interest is ChOPS J182841.47-040535.6, which has an early-M type profile with both
Balmer and \ion{He}{1} $\lambda$6678 \AA\ and $\lambda$5875 \AA\ emission lines.  The \ha\ line is extremely
strong, with equivalent width $\sim$90 \AA.  Based on similarities to flare star emission lines reported
by \citet*{giz2002}, this object is most likely a flaring M star.  As it fell outside the \emph{Chandra}
field of view, we have no X-ray data for this object.  
A full catalog of these objects, with equivalent width
determinations and dwarf/giant differentiation, will be
forthcoming in later publications of the project.  

\subsection{Quasars \& AGN}
\begin{figure}
\plotone{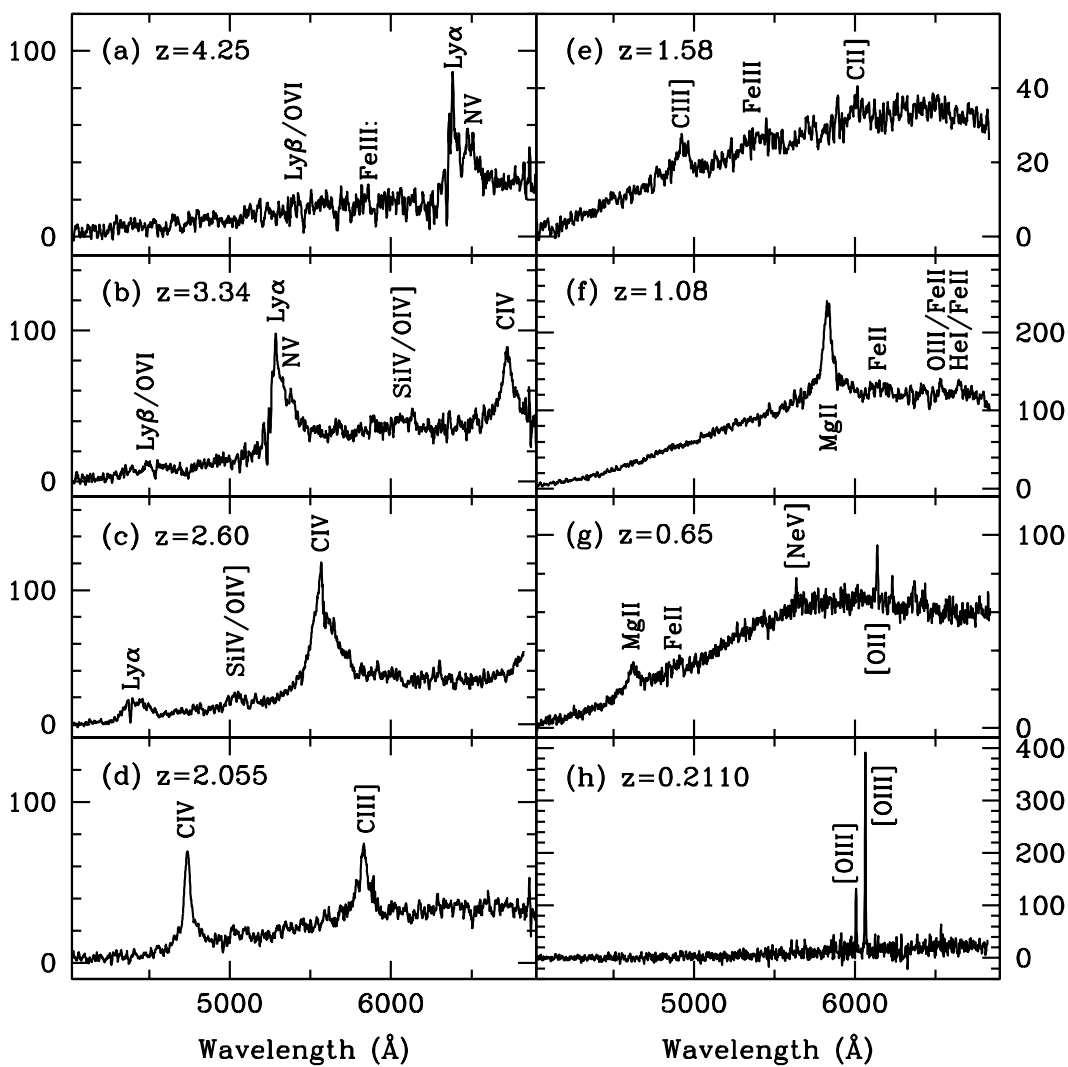}
\caption{Representative quasar spectra from varous fields in the
ChaMPlane project.  Most are regular broad-line objects, while (h) is
a Seyfert II.  \label{quasar}}
\end{figure}

A final serendipitous result of the ChaMPlane project is the discovery of 29
quasars or lower luminosity active galactic nuclei (AGN), 24 of which were detected
as X-ray emission objects.  
Specific data on each quasar detected during
the project to date are given in Table \ref{tbl-quasars}, with
all CXOPS quasars being X-ray detections.  
While not ideal for quasar detection owing to obscuration in
the Galactic plane, the ChaMPlane project has nevertheless been quite 
successful in detecting these objects.  Quasars with redshifts ranging
from 0.02 to 4.25 have been observed, with representative spectra shown
in Figure \ref{quasar}.  The classification of a spectrum as a quasar is conservatively based
on being able to find a minimum of two emission features corresponding to lines in the
composite quasar spectrum of \citet{vanb2001}.  
Most quasars are detected as X-ray sources, with five being detected as apparent \ha\ sources.
Two of these '\ha' objects are narrow-line AGN, possibly Seyfert IIs \citep[like 
Figure \ref{quasar} panel (h), by comparison
with Seyfert II spectra in][]{mor2002},
and three are high-redshift broad-line 
objects.  Of these three, two lie within the \emph{Chandra} field of view but were not
detected as X-ray sources, thus they must be weak X-ray emitters.  The narrow-line AGNs and the other
broad-line AGN lie outside the \emph{Chandra} field, so we have no X-ray data on these three sources.
In all cases, the `\ha' detection is 
actually another line redshifted
into the narrow wavelength range of the \ha\ filter.  
Our WIYN/Hydra spectroscopic program is 
limited in quasar detection to those with redshifts less 
than about 4.6, owing to
the shift of Ly$\alpha$ past our red spectral limit of 
6800 \AA\ at higher redshift.  Thus, quasars
at higher redshifts would have 
a very low, nearly featureless continuum present in our spectra, 
and would not be able to be classified.
Fainter quasars 
at a redshift around 1.0 will also not be detected by our spectroscopy, as there
is only one major emission feature (\ion{Mg}{2} $\lambda$2798 \AA) available.  Without 
secondary lines, such as in Figure \ref{quasar} panel (f), a positive quasar 
classification with redshift determination cannot be made.   

Examination of our quasar data reveals that 16 of 29 quasars are located in a single
field.  This field, J0422+32, has the lowest average E(B--V) of any field in the
survey.  Additionally, examination of the spectra shows that this interstellar gas is
concentrated in a portion of the field, leaving the rest of the field with extremely low
interstellar extinction as compared to other fields.  
This is also seen in the extinction map of the region based on data from 
\citet{schl1998}.   
The Chandra Multiwavelength Project (ChaMP), 
which is an extragalactic serendipitous \emph{Chandra} source AGN survey, 
reports an average of
10 AGN per \emph{Chandra} field, with 12 fields surveyed \citep{silv2003}.  For our J0422 field, 11 of 
the 16 quasars are detected by X-ray emission, and 
two more lie within the \emph{Chandra} field of view.
Therefore, the J0422 field is similar in quasar detection rates
to non-Galactic Plane fields, supporting the inference of extremely low interstellar extinction
in most of this field.  
A detailed investigation of the ChaMPlane AGN, comparing their X-ray 
and optical properties, will be published separately.

\section{Discussion and Conclusions}

On examination of the overall data set, some trends are evident in the object
classifications.
All of our five CV candidates
are located in the 80\% of each field imaged only with the Mosaic camera,
and thus only show up on the \ha\ lists.  The two known LMXBs
showed up, as expected, on both the \ha\ and X-ray target lists.  
The Be stars discovered so
far are all at the very bright end of our target objects, 
owing to the concentration of star forming activity within the solar circle.

The five new CVs and two known LMXBs 
are distributed across the surveyed magnitude range.
Most quasars are optically faint X-ray objects, as would be expected from the strong increase 
of quasar counts with redshift.
Coronal emission stars make up the bulk of the 
brighter X-ray targets, 
with some nearby normal (non-emission) stars at the
very brightest end of our magnitude range.  Both \ha\ lists contain primarily
late-type stars, 
with about $45-50\%$ (in the \ha-definite list, with \har$<-0.3$) and about 
$30-45\%$ (in the \ha-marginal list, with $-0.3<$\har$<-0.2$)
 appearing as coronal emission objects 
regardless of magnitude.  
The predominance of late-type stars in our \ha-selected sample is not surprising,
given the locations of the strong TiO absorption bands in the continuum of 
the M-star spectrum at about \ha\ causing them to have an \har\ color closer to our threshold
value than earlier-type stars.  

In summary, during the first three years of followup spectroscopy of northern fields 
for the ChaMPlane
project, we have found five new CV candidates, 
four new Be star candidates, 14 lithium-absorption stars, 29 new quasars,
and many new coronal emission objects.  While at this time the total number
of CVs detected is small, we have shown that 
our program of initial Mosaic photometry and followup WIYN-Hydra 
spectroscopy detects CVs in a systematic fashion.  The larger samples expected
from the full ChaMPlane survey will allow a 
a volume-limited sample determination of the CV space density.  Our 
serendipitous discoveries of Li stars, AGN, and substantial numbers of coronal emission
stars will all lend themselves to extensive further investigation.

\acknowledgments
This research was supported in part by NSF grant AST-0098683, 
NASA/Chandra grants AR2-3002B, AR3-4002B, and AR4-5003B.  ABR 
gratefully acknowledges a fellowship from the Indiana Space Grant Consortium.

\clearpage

\clearpage
\begin{deluxetable}{llrrrc}
\tablecaption{Fields observed Jan 2001-Feb 2004.  \label{tbl-fielddata}}
\tablewidth{0pt}
\tablehead{
\colhead{Field} & \colhead{RA} & \colhead{DEC} & \colhead{l} & \colhead{b} & 
\colhead{$E(B-V)$}
}
\startdata
GK Persei &03:31.2 &+43:54 &151.0 & $-$10.1 & 0.4\\
GRO J0422+32 &04:21.7 &+32:54 &165.9 & $-$11.9 & 0.3\\
NGC 1569 &04:30.8 &+64:50 &143.7 & 11.2 & 0.7\\
3C 123 &04:36.4 &+29:38 &170.5 &$-$11.8 & 1.0\\
3C 129 &04:49.3 &+45:03 &160.4 &0.2 & 1.1\\
PSRJ0538+2817 &05:38.0 &+28:14 &179.7 &$-$1.7 & 1.4\\
A0620-00 &06:23.3 &$-$00:18 &210.0 &$-$6.5 &0.5\\
M1-16 &07:37.3 &$-$09:39 &226.8 &5.6 &0.2\\
MWC297 &18:27.7 &$-$03:50 &26.8 &3.5 &20.4\\
SGR 1900+14 &19:07.2 &+09:19 &43.0 &0.8 &4.2\\
B2224+65 &22:25.9 &+65:36 &108.6 &6.8 &0.7\\
3EG J2227+6122 &22:29.3 &+61:19 &106.7 &3.0 & 1.7\\
G116.9+0.2 &23:59.2 &+62:24 &116.9 &0.2 &0.8\\
\enddata
\end{deluxetable}

\clearpage
\begin{deluxetable}{llr}
\tablecaption{Summary of ChaMPlane-WIYN Observations.  \label{tbl-fieldepochs}}
\tablewidth{0pt}
\tablehead{
\colhead{Epoch} &\colhead{Field} &\colhead{Total Exp.(min.)}
}
\startdata
2001 Jan 23--25 & J0422 & 300 \\
\nodata & A0620 & 300 \\
2002 Jan 07--10 & GK Per & 420 \\
\nodata & J0422 & 280 \\
\nodata & NGC 1569 & 360 \\
\nodata & 3C 123 & 420 \\
2002 Aug 3-7 & B2224 & 354\\
\nodata &J2227 &399\\
2002 Oct 2 & 3C 129 & 345\\
\nodata & SGR 1900 & 180\\
2002 Oct 31 & A0620 & 220 \\
\nodata & B2224 & 200\\
2002 Dec 5 & A0620 &235\\
\nodata & B2224 & 200\\
2003 Jun 23--25 & MWC297 & 440\\
\nodata & SGR1900 & 360\\
\nodata & J2227 &170\\
2003 Sept 20--24 & PSRJ0538 & 237\\
\nodata & G116 & 340\\
2004 Feb 21--23 & M1-16 & 60
\enddata
\end{deluxetable}

\clearpage
\begin{deluxetable}{lrrr}
\tablecaption{Observation summary by target type.  \label{tbl-obssum}}
\tablewidth{0pt}
\tablehead{
\colhead{Target Type} & \colhead{Total Targets} & \colhead{Number Observed} & \colhead {Fraction Classified}}
\startdata
X-ray &363 &262 &48$\%$\\
H$\alpha$ &1053 &273 &51$\%$\\
H$\alpha$M &1090 &302 &81$\%$\\
Other &565 &224 &49$\%$\\
Duplicate\tablenotemark{a} &14 &14 &86$\%$\\
\enddata
\tablenotetext{a}{These objects appear both on the X-ray list and one of the H$\alpha$ lists.}
\end{deluxetable}

\clearpage
\begin{deluxetable}{lllrrcll}
\tabletypesize{\scriptsize}
\tablecaption{All CV and LMXB candidates observed to date. \label{tbl-cvs}}
\tablewidth{0pt}
\tablehead{
\colhead{Ref. \#} &\colhead{Field} &
\colhead{Name} & \colhead{RA} & 
\colhead{DEC} & \colhead{R mag.} & 
\colhead{XFOV\tablenotemark{a}} &
\colhead{notes}}
\startdata
XB 1 &A0620 &A0620-00 &06 22 44.56 &$-$00 20 44.3 &17.37 &Y\tablenotemark{b}& \tablenotemark{c} \\
XB 2 &J0422 &J0422+32 &04 21 42.71 &32 54 27.1 &20.83 &Y\tablenotemark{b}& \tablenotemark{d} \\
CV 1 &MWC297 &ChOPS J182723.91-040715.9 &18 27 23.92 &$-$04 07 15.9 &19.33 &N& \tablenotemark{e} \\
CV 2 &G116   &ChOPS J235902.60+621325.1 &23 59 02.61 &62 13 25.1 &19.71 &N& \tablenotemark{f} \\
CV 3 &J0422 &ChOPS J042130.28+330729.2 &04 21 30.28 &33 07 29.2 &20.33 &N& \nodata \\
CV 4 &3C 123 &ChOPS J043534.31+292037.9 &04 35 34.32 &29 20 37.9 &21.65 &N&\nodata \\
CV 5 &M1-16 &ChOPS J073810.09-092341.1 &07 38 10.10 &$-$09 23 41.2 &21.58 &N&\nodata \\
\enddata
\tablenotetext{a}{Indicates whether the object is located in the \emph{Chandra} ACIS field of view.}
\tablenotetext{b}{This object was detected by \emph{Chandra}.}
\tablenotetext{c}{This object is known to be a LMXB with black hole primary \citep{a0620}.}
\tablenotetext{d}{This object is known to be a LMXB with black hole primary \citep{j0422}.}
\tablenotetext{e}{This object has a definite M-star continuum present.}
\tablenotetext{f}{This object has \ion{He}{2} emission at $\lambda$4686 \AA.}
\end{deluxetable}

\clearpage

\begin{deluxetable}{llrrccl}
\tabletypesize{\footnotesize}
\tablecaption{New Be star candidates.  \label{tbl-bes}}
\tablewidth{0pt}
\tablehead{
\colhead{Field} & \colhead{Name} & \colhead{RA} & \colhead{DEC} & 
\colhead{R mag.} & \colhead{XFOV\tablenotemark{a}} &\colhead{Class}}
\startdata
3C 129 &ChOPS J044927.22+450443.8 &04 49 27.22 &45 04 43.8 &13.61 & Y&~B5\\
J2227  &ChOPS J223048.26+611453.7 &22 30 48.26 &61 14 52.7 &15.13 & N&late O\\
J2227  &ChOPS J223030.43+612255.3 &22 30 30.43 &61 22 55.3 &16.52 & Y&early B\\
J2227  &ChOPS J223013.39+613140.8 &22 30 13.39 &61 31 40.8 &18.83 & N&B\\
\enddata
\tablenotetext{a}{Indicates whether the object is located in the \emph{Chandra} ACIS field of view.
None of these objects were detected by \emph{Chandra}.}
\end{deluxetable}

\begin{deluxetable}{llrrcll}
\tabletypesize{\footnotesize}
\tablecaption{Li stars. \label{tbl-li}}
\tablewidth{0pt}
\tablehead{
\colhead{Field} &
\colhead{Name} &
\colhead{RA} &
\colhead{DEC} &
\colhead{R mag.} &
\colhead{XFOV\tablenotemark{a}} &
\colhead{EW\tablenotemark{b}}
}
\startdata
MWC297 &ChOPS J182737.07-034937.9&18 27 37.07&$-03$\ 49 38.0&17.13&Y& 0.4\\
GKPer  &CXOPS J033127.7+434623   &03 31 27.60&43 46 22.4&16.51&Y\tablenotemark{c}& 0.4\\
NGC1569&CXOPS J043003.1+645142   &04 30 03.18&64 51 42.6&17.88&Y\tablenotemark{c}& 0.4\\
J2227  &CXOPS J222833.5+611105   &22 28 33.51&61 11 05.7&15.09&Y\tablenotemark{c}& 0.3\\
J2227  &ChOPS J222802.93+610430.1&22 28 02.94&61 04 30.2&16.86&N & 0.5\\
SGR1900&CXOPS J190703.1+092215   &19 07 03.23&09 22 14.5&17.14&Y\tablenotemark{c}& 0.3\\
B2224  &ChOPS J222729.04+654026.2&22 27 29.04&65 40 26.3&17.86&N & 0.4\\
B2224  &CXOPS J222413.3+653628   &22 24 13.67&65 36 26.4&17.04&Y\tablenotemark{c}& 0.4\\
B2224  &ChOPS J222632.12+654954.7&22 26 32.12&65 49 54.8&16.18&N & 0.6\\
A0620  &ChOPS J062222.14-002515.2&06 22 22.15&$-00$\ 25 15.3&19.11&N& 0.4\\ 
A0620  &CXOPS J062159.6-001458   &06 21 59.62&$-00$\ 14 56.2&\nodata\tablenotemark{d}&Y\tablenotemark{c}& 0.2\\
MWC297 &ChOPS J182649.73-035559.1&18 26 49.73&$-03$\ 55 59.2&17.71&N & 0.6\\
J0538  &ChOPS J053828.61+282845.8&05 38 28.61&28 28 45.9&18.76&Y &0.6\\
J0538  &CXOPS J053831.5+281459   &05 38 31.55&28 14 59.4&17.12&Y\tablenotemark{c}&0.6\\
A0620  &A0620-00                 &06 22 44.56&$-00$\ 20 44.3&17.37& Y\tablenotemark{c}&0.2\tablenotemark{e}\\
\enddata
\tablenotetext{a}{Indicates whether the object is located in the \emph{Chandra} ACIS field of view.}
\tablenotetext{b}{Equivalent width of the Lithium $\lambda 6708$\AA\ absorption line.}
\tablenotetext{c}{This object was detected by \emph{Chandra}.}
\tablenotetext{d}{This target was off the Mosaic field of view.}
\tablenotetext{e}{A0620-00 is known to have high lithium abundance\citep{gon2004}.}
\end{deluxetable}

\clearpage
\begin{deluxetable}{llrrclcl}
\tabletypesize{\scriptsize}
\tablecaption{Quasar candidates detected to date by ChaMPlane/WIYN.
\label{tbl-quasars}}
\tablewidth{0pt}
\tablehead{
\colhead{Field} & \colhead{Name} & \colhead{RA\tablenotemark{a}} & 
\colhead{DEC\tablenotemark{a}} & \colhead{R mag.} & \colhead{z} &
\colhead {Conf.\tablenotemark{b}} & \colhead{Notes}
}
\startdata
NGC1569 &CXOPS J043041.4+643925    &04 30 41.21 &64 39 25.6 &21.24 &0.2110 &5 &\tablenotemark{c,d}\\	
NGC1569 &CXOPS J043125.1+645154    &04 31 25.13 &64 51 54.3 &20.04 &0.279 &5 &\tablenotemark{c,e}\\	
J0422   &ChOPS J042048.59+324521.9 &04 20 48.59 &32 45 22.0 &22.71 &0.3070 &5 &\tablenotemark{c,f}\\	
J0422   &ChOPS J042051.71+324533.8 &04 20 51.72 &32 45 33.8 &23.50 &0.3099 &5 &\tablenotemark{c,f}\\	
J0422   &CXOPS J042211.8+325605    &04 22 11.83 &32 56 04.3 &20.39 &0.65 &4 & \nodata \\	
GKPer   &CXOPS J033105.9+440328    &03 31 05.95 &44 03 27.6 &20.41 &0.93 &5 & \nodata \\	
GKPer   &CXOPS J033136.5+434209    &03 31 36.66 &43 42 11.3 &19.96 &1.08 &5 &\tablenotemark{h}\\
J0422   &CXOPS J042155.0+330037    &04 21 55.04 &33 00 36.3 &21.03 &1.125 &4 & \nodata \\
B2224   &CXOPS J222456.1+653131    &22 24 56.21 &65 31 31.5 &22.27 &1.13 &2 & \nodata \\
G116    &CXOPS J235813.4+622447    &23 58 13.47 &62 24 47.8 &21.57 &1.29 &4 & \nodata \\
3C123   &CXOPS J043657.6+294155    &04 36 57.63 &29 41 55.5 &21.28 &1.30 &3 & \nodata \\
J0422   &CXOPS J042127.7+325038    &04 21 27.76 &32 50 38.3 &22.07 &1.31 &3 & \nodata \\
J0422   &ChOPS J042140.57+331204.1 &04 21 40.58 &33 12 04.1 &21.15 &1.35 &2 & \tablenotemark{f} \\
J0422   &ChOPS J042028.32+325239.7 &04 20 28.32 &32 52 39.8 &22.44 &1.40 &3 & \tablenotemark{g} \\
J0422   &CXOPS J042201.0+325237    &04 22 00.97 &32 52 36.4 &20.92 &1.58 &3 & \nodata \\
J0422   &CXOPS J042155.1+324726    &04 21 55.08 &32 47 26.0 &20.98 &1.845 &5 & \nodata \\
GKPer   &CXOPS J033112.2+433844    &03 31 12.16 &43 38 44.4 &20.70 &1.85 &4 & \nodata \\
J0422   &CXOPS J042117.5+330206    &04 21 17.46 &33 02 05.5 &19.23 &1.93 &3 & \nodata \\
J0422   &CXOPS J042140.7+324941    &04 21 40.83 &32 49 40.7 &22.82 &1.94 &4 & \nodata \\
G116    &CXOPS J235813.2+623343    &23 58 13.29 &62 33 44.6 &21.11 &2.04 &5 & \nodata \\
J0422   &CXOPS J042133.8+325557    &04 21 33.84 &32 55 57.7 &21.07 &2.055 &5 & \nodata \\
A0620   &CXOPS J062240.4-002113    &06 22 40.47 &$-$00 21 13.6 &22.65 &2.20 &5 & \nodata \\
J0422   &CXOPS J042201.6+325730    &04 22 01.58 &32 57 29.3 &21.56 &2.203 &3 & \nodata \\
GKPer   &CXOPS J033120.7+434002    &03 31 20.64 &43 40 01.2 &21.56 &$2.60\pm0.03$ &4 &\nodata \\
J0422   &CXOPS J042154.3+325310    &04 21 54.39 &32 53 09.8 &21.55 &2.91 &5 & \nodata \\
B2224   &CXOPS J222525.0+652917    &22 25 25.17 &65 29 17.7 &22.31 &3.06 &3 & \nodata \\
M1-16   &CXOPS J073749.6-094432    &07 37 49.73 &$-$09 44 31.7 &23.05 &3.18 &3 &\nodata \\
J0422   &CXOPS J042210.3+330126    &04 22 10.35 &33 01 26.0 &21.12 &3.34 &5 & \nodata \\
J0422   &ChOPS J042200.24+325708.0 &04 22 00.25 &32 57 08.0 &22.01 &4.25 &3 & \tablenotemark{g} \\
\enddata
\tablenotetext{a}{RA and DEC are taken from the optical counterpart positions.}

\tablenotetext{b}{Confidence of redshift determination (based on line identifications): 5=highest, 2=low.  Possible quasars with
confidence of 1 were not included in the table.}

\tablenotetext{c}{This object appears to be a narrow-line AGN.}

\tablenotetext{d}{This object is likely a Seyfert II, by comparison with spectra in \citet{mor2002}.}

\tablenotetext{e}{This object is in the literature \citep{CXOU} as 
	CXOU J043125.1+645154 as
an AGN.}

\tablenotetext{f}{This object was not in the \emph{Chandra} field of view.}

\tablenotetext{g}{This object was in the \emph{Chandra} field of view but was not detected.}

\tablenotetext{h}{This object is in the literature \citep{1RXH} as 
	1RXH J033136.5+434213 as
an X-ray source.}

\end{deluxetable}


\clearpage

\begin{thebibliography}{}

\bibitem[Binney \& Merrifield(1998)]{bin1998} Binney, J. \& Merrifield, M. 1998, Galactic Astronomy (Princeton: Princeton University Press)

\bibitem[Callanan, et al.(1996)]{j0422} Callanan, P. J., Garcia, M. R., McClintock, J. E., Zhao, P., Remillard, R. A., \& Haberl, F. 1996, \apj, 461, 351

\bibitem[Drake, et al.(2002)]{dra2002} Drake, N. A., de la Reza, R., da Silva, L., Lambert, D. L. 2002, \aj, 123, 2703, astro-ph/2020158

\bibitem[ESA(1997)]{hipp1997} ESA 1997, The Hipparcos and Tycho Catalogues, ESA SP-1200

\bibitem[Gizis et al.(2002)Gizis, Reid, \& Hawley]{giz2002} Gizis, J. E., Reid, I. N., \& Hawley, S. L. 2002, \aj, 123, 3356, astro-ph/0203499

\bibitem[Gonz\'{a}lez Hern\'{a}ndez, et al.(2004)]{gon2004} Gonz\'{a}lez Hern\'{a}ndex, J. I., Rebolo, R., Israelian, G., Casares, J., Maeder, A., \& Meynet, G. 2004, \apj, 609, 988, astro-ph/0403402

\bibitem[Grindlay et al.(2003)]{gri2003} Grindlay, J. E., et al. 2003, Astron. Nachr., 324, 57, astro-ph/0211527

\bibitem[Grindlay et al.(2005)]{gri2004} Grindlay, J. E., et al. 2005, in preparation

\bibitem[Haswell et al.(1993)]{a0620} Haswell, C. A. Robinson, E. L., Horne, K., Rae, F., \& Abbott, T. M. C. 1993, \apj, 411, 802

\bibitem[Hatzidimitriou et al.(2003)]{hat2003} Hatzidimitriou, D., Morgan, D. H., Cannon, R. D., \& Croke, B. F. W. 2003, \mnras, 341, 1290, astro-ph/0304297

\bibitem[Hong et al.(2005)]{hon2004} Hong, et al. 2005, in preparation

\bibitem[K\"{u}pc\"{u} Yoldas \& Balman(2002)]{1RXH} K\"{u}pc\"{u} Yoldas, A. \& Balman, S. 2002, \aap, 384, 190, astro-ph/0201111

\bibitem[Laycock et al.(2005)]{lay2004} Laycock, et al. 2005, in preparation

\bibitem[Martin, Kobulnicky, \& Heckman(2002)]{CXOU} Martin, C. L., Kobulnicky, H. A., \& Heckman, T. M. 2002, \apj, 574, 663, astro-ph/0203513

\bibitem[Moran, Filippenko, \& Ryan(2002)]{mor2002} Moran, E. C., Filippenko, A. V., \& Ryan, C. 2002, \apj, 579, L71

\bibitem[Porter \& Rivinius(2003)]{por2003} Porter, J. M. \& Rivinius, T. 2003, \pasp, 115, 1153

\bibitem[Reid \& Hawley(2000)]{rei2000} Reid, I. N. \& Hawley, S. L. 2000, New Light on Dark Stars: Red Dwarfs, Low-mass Stars, Brown Dwarfs (Berlin: Springer-Verlag) 

\bibitem[Rogel et al.(2005)]{rog2004} Rogel, A., et al. 2005, in preparation

\bibitem[Silverman et al.(2003)]{silv2003} Silverman, J., et al. 2003, Astron. Nachr., 324, 97

\bibitem[Schlegel, et al.(1998)]{schl1998} Schlegel, D. J., Finkbeiner, D. P., \& Davis, M. 1998, \apj, 599, 525

\bibitem[Smith et al.(1999)]{smi1999} Smith, K.W., Lewis, G.F., Bonnell, I.A., Bunclark, P.S., \& Emerson, J.P. 1999, \mnras, 304, 367

\bibitem[Steele et al.(1999)Steele, Negueruela, \& Clark]{ste1999} Steele, I. A., Negueruela, I., \& Clark, J. S. 1999, \aaps, 137, 147, astro-ph/9906245

\bibitem[The, Steenman, \& Alcaino(1984)]{thep1984} The, P. S., Steenman, H. C., \& Alcaino, G. 1984, \aap, 132, 385

\bibitem[Vanden Berk et al.(2001)]{vanb2001} Vanden Berk, D., et al. 2001 \aj, 122, 549

\bibitem[Winkler et al.(1997)]{wink1997} Winkler, H., 1997, \mnras, 287, 481 

\bibitem[Zhao et al.(2003)]{zha2003} Zhao, P., Grindlay, J., Edmonds, P., Hong, J., Jenkins, J., Schlegel, E., Cohn, H., \& Lugger, P.  2003, Astron. Nachr., 324, 176, astro-ph/0405509

\bibitem[Zhao et al.(2005)]{zha2004} Zhao, P., et al. 2005, in preparation

\end{thebibliography}
\end{document}